\documentclass[pre,twocolumn,amsmath,amssymb,floatfix,superscriptaddress,nopacs]{revtex4}

\usepackage{graphicx}
\usepackage{latexsym}
\usepackage{amsmath}
\usepackage{amssymb}
\usepackage{amsfonts}
\usepackage{bm}

\newcommand{\ddiff}{\ensuremath{\mathrm{d}}}

\newcommand{\Tglass}{\mbox{$T_{\rm g}$}}

\newcommand{\muA}{\ensuremath{\mu_\mathrm{A}}}
\newcommand{\dmuA}{\ensuremath{\delta \mu_\mathrm{A}}}
\newcommand{\muAbar}{\ensuremath{\overline{\mu}_\mathrm{A}}}
\newcommand{\muAhat}{\ensuremath{\hat{\mu}_\mathrm{A}}}
\newcommand{\muF}{\ensuremath{\mu_\mathrm{F}}}
\newcommand{\dmuF}{\ensuremath{\delta \mu_\mathrm{F}}}
\newcommand{\muFbar}{\ensuremath{\overline{\mu}_\mathrm{F}}}
\newcommand{\muFone}{\ensuremath{\mu_1}}
\newcommand{\dmuFone}{\ensuremath{\delta \mu_1}}
\newcommand{\muFonebar}{\ensuremath{\overline{\mu}_1}}
\newcommand{\muFtwo}{\ensuremath{\mu_0}}
\newcommand{\dmuFtwo}{\ensuremath{\delta \mu_0}}
\newcommand{\muFtwobar}{\ensuremath{\overline{\mu}_0}}
\newcommand{\Geq}{\mu}
\newcommand{\GFbar}{\overline{\mu}}
\newcommand{\GF}{\mu}
\newcommand{\dGF}{\delta \mu}
\newcommand{\GFmed}{\mu_\mathrm{med}}
\newcommand{\GFmax}{\mu_\mathrm{max}}

\newcommand{\tauhat}{\ensuremath{\hat{\tau}}}

\newcommand{\tsamp}{\Delta t}
\newcommand{\tsampmax}{\Delta t_\mathrm{max}}

\begin{document}

\title{Shear modulus and shear-stress fluctuations in polymer glasses}

\author{I. Kriuchevskyi}
\affiliation{Institut Charles Sadron, Universit\'e de Strasbourg \& CNRS, 23 rue du Loess, 67034 Strasbourg Cedex, France}
\author{J.P.~Wittmer}
\email{joachim.wittmer@ics-cnrs.unistra.fr}
\affiliation{Institut Charles Sadron, Universit\'e de Strasbourg \& CNRS, 23 rue du Loess, 67034 Strasbourg Cedex, France}
\author{H. Meyer}
\affiliation{Institut Charles Sadron, Universit\'e de Strasbourg \& CNRS, 23 rue du Loess, 67034 Strasbourg Cedex, France}
\author{J. Baschnagel}
\affiliation{Institut Charles Sadron, Universit\'e de Strasbourg \& CNRS, 23 rue du Loess, 67034 Strasbourg Cedex, France}

\begin{abstract}
Using molecular dynamics simulation of a standard coarse-grained polymer glass model 
we investigate by means of the stress-fluctuation formalism the shear modulus $\GF$ 
as a function of temperature $T$ and sampling time $\tsamp$.
While the ensemble-averaged modulus $\GF(T)$ is found to decrease continuously for all $\tsamp$ sampled,
its standard deviation $\dGF(T)$ is non-monotonous with a striking peak at the glass transition.
Confirming the effective time-translational invariance of our systems, $\GF(\tsamp)$ can be understood
using a weighted integral over the shear-stress relaxation modulus $G(t)$.
While the crossover of $\GF(T)$ gets sharper with increasing $\tsamp$,
the peak of $\dGF(T)$ becomes more singular.
It is thus elusive to predict the modulus of a single configuration at the glass transition.
\end{abstract}
\date{\today}
\maketitle

\paragraph*{Introduction.}
The shear modulus $\Geq$ is the central, mechanically directly accessible, 
order parameter characterizing the transition from the liquid/sol ($\Geq=0$)
to the solid/gel state ($\Geq > 0)$
\cite{LandauElasticity,DoiEdwardsBook,HansenBook,RubinsteinBook,Alexander98}.
Since the shear modulus $\Geq(T)$ of crystalline solids vanishes discontinuously at the melting point
with increasing temperature $T$ \cite{LXW16}, this begs the question of the behavior of $\Geq(T)$ for
amorphous solids near the glass transition temperature $\Tglass$ 
\cite{Barrat88,Yoshino12,ZT13,WXP13,LXW16,GoetzeBook,Szamel11,Ikeda12,Klix12,Klix15,Yoshino14}.
Two qualitatively different scenarios have been put forward, being either in favor of
a {\em continuous} (cusp-like) transition \cite{Barrat88,Yoshino12,ZT13,WXP13,LXW16} or of
a {\em discontinuous} jump at $\Tglass$
\cite{GoetzeBook,Szamel11,Ikeda12,Klix12,Klix15,Yoshino14}.
The jump singularity is a result of mean-field theories \cite{GoetzeBook,Biroli09,Yoshino14} which 
find the energy barriers for complete structural relaxation to diverge at $\Tglass$ so that liquid-like 
flow stops. However, in experimental or simulated glass formers the barriers do not diverge abruptly. 
Such non-mean-field effects are expected to smear out the sharp transition \cite{Yoshino14}.  
Another line of recent research has focused on the elastic properties deep in the glass
\cite{Gardner,Biroli16,Procaccia16}. 
At $T \ll \Tglass$ a transition in the solid is found, where multiple particle arrangements occur 
as different competing glassy states. This so-called ``Gardener" transition is thus accompanied by 
strong fluctuations of $\Geq$ from one glass state to the other \cite{Biroli16,Procaccia16}.
Interestingly, strong fluctuations of $\Geq$ were also observed in amorphous self-assembled networks 
\cite{WKC16} (a model for vitrimers \cite{Leibler11}). 
The results of \cite{Biroli16,Procaccia16,WKC16} beg the question of whether the emergence of shear 
rigidity at the glass transition is also accompanied by strong fluctuations of $\Geq$.
Here we address both questions by means of large-scale molecular dynamics (MD) 
\cite{AllenTildesleyBook} 
simulations of a standard model for glassy polymers 
\cite{LAMMPS,Frey15,BaschRev16,SBM11,ivan17a}. 
Details about the model, quench protocol and measured observables 
may be found in the Supplemental Material (SM) \cite{foot_SM}.
Lennard-Jones units \cite{AllenTildesleyBook} are used below.

\begin{figure}[t]
\centerline{\resizebox{.9\columnwidth}{!}{\includegraphics*{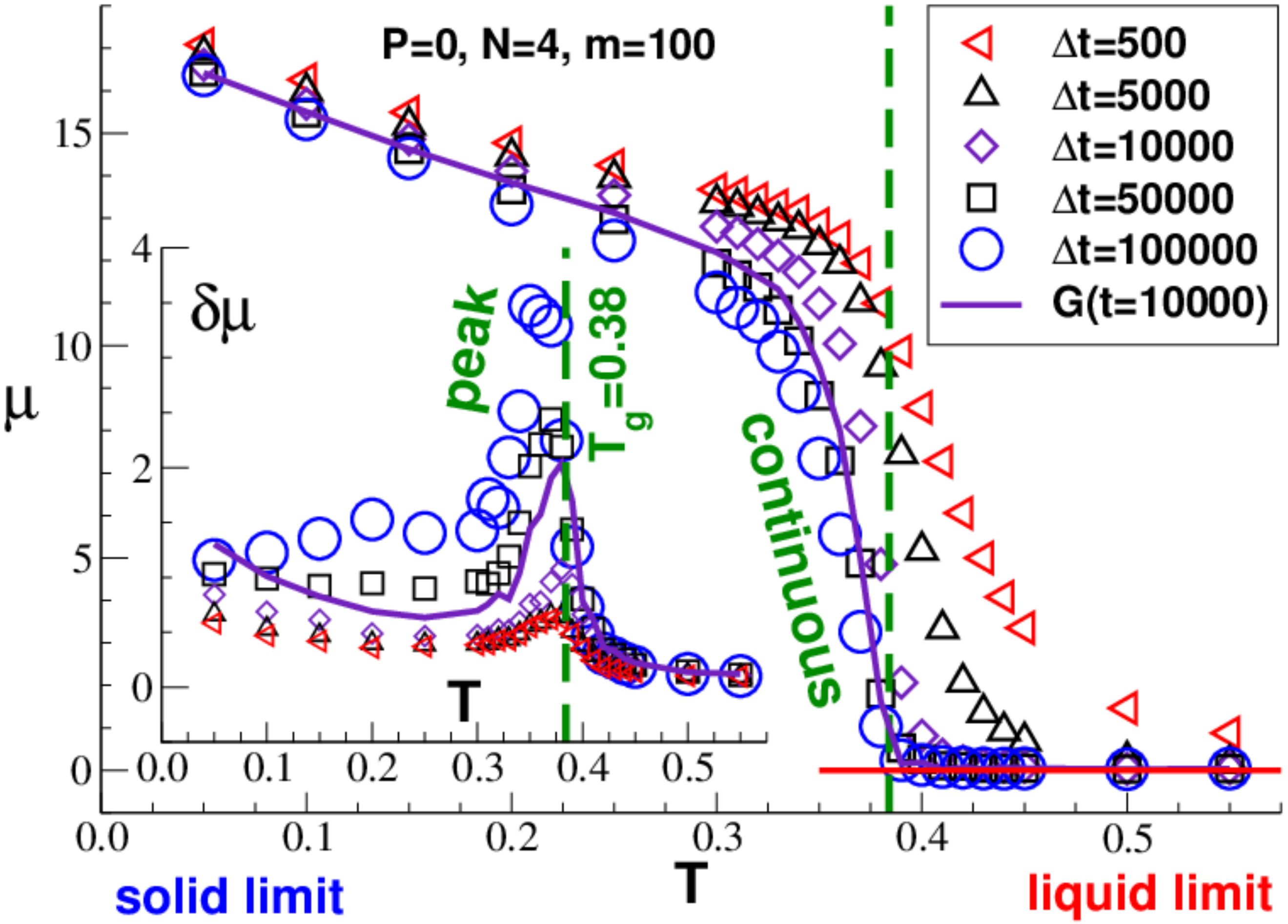}}}
\caption{Key findings as a function of temperature $T$. 
The vertical dashed lines indicate the glass transition temperature $\Tglass \approx 0.38$.
Main panel:
Shear modulus $\GF(T)$ for different sampling times $\tsamp$ showing that the transition
becomes more and more step-like with increasing $\tsamp$.
Inset:
Corresponding standard deviation $\dGF(T)$ showing a peak at $T \approx \Tglass$
which becomes sharper with increasing $\tsamp$.
Also included are the shear stress relaxation modulus $G(t)$ and its standard deviation
$\delta G(t)$ taken at a time $t=10^4$ (bold solid lines).
}
\label{fig_key}
\end{figure}

\paragraph*{Key findings.}
Following the pioneering work by Barrat {\em et al.} \cite{Barrat88}
and many recent numerical studies 
\cite{WTBL02,Barrat06,SBM11,XWP12,WXP13,LXW16,Procaccia16,ivan17a} 
we use the stress-fluctuation formalism \cite{Hoover69,Lutsko88,WXB15,WXBB15,WKB15} 
to determine the shear modulus.
Our key findings for $\GF$ and its standard deviation $\dGF$, obtained as function of $T$ for 
a broad range of sampling times $\tsamp$, are summarized in Fig.~\ref{fig_key}. 
Albeit $\GF(T)$ remains always continuous, it becomes systematically more step-like with increasing $\tsamp$. 
At variance to the monotonous modulus $\GF(T)$ its standard deviation $\dGF(T)$ is non-monotonous 
with a remarkable peak near $\Tglass \approx 0.38$. 
(As explained in the SM \cite{foot_SM}, $\Tglass$ is defined here by means of a 
$\tsamp$-independent dilatometric criterion during the initial
continuous temperature quench \cite{SBM11,LXW16}.) 
The peak increases with $\tsamp$, becoming about a third of the drop of $\GF(T)$ between $T=0.34$ and $T=0.38$
for $\tsamp=10^5$. The liquid-solid transition is thus accompanied by strong fluctuations between
different glass configurations.
We corroborate these results below.
Especially, we shall trace back the observed $\tsamp$-dependence of $\GF$ to the 
time dependence of the shear-stress relaxation modulus $G(t)$ \cite{RubinsteinBook}. 

\paragraph*{Time series.}
Our $m=100$ independently quenched configurations contain $3072$ chains of length $N=4$.
A vanishing normal pressure ($P=0)$ is imposed for all $T$.
Having reached a specific temperature and after tempering over $\tsampmax=10^5$
we perform production runs over again $\tsampmax$ with entries made each velocity-Verlet sweep.
Of importance are here the instantaneous shear stress $\tauhat$ and the instantaneous ``affine
shear modulus" $\muAhat$.
As reminded in Sec.~2 of the SM \cite{foot_SM}, $\tauhat$ is the first functional derivative
of the Hamiltonian with respect to an imposed infinitesimal canonical
and affine shear transformation and $\muAhat$ the corresponding second functional derivative
\cite{WXP13,WXB15,WXBB15,WKB15,WXB16,WKC16}. 
The stored time-series are used to compute for a given configuration and shear plane
various {\em time-averages} \cite{AllenTildesleyBook} (marked by horizontal bars)
over sampling times $\tsamp \le \tsampmax$
\begin{eqnarray}
\muAbar    & \equiv & \overline{\muAhat} \label{eq_muAbar} \\
\muFbar    & \equiv & \muFtwobar - \muFonebar      \label{eq_muFbar}
\mbox{ with } \muFtwobar \equiv \beta V \overline{\tauhat^2}, \
              \muFonebar \equiv \beta V \overline{\tauhat}^2 \\
\GFbar     & \equiv & \muAbar - \muFbar \equiv (\muAbar-\muFtwobar) + \muFonebar  \label{eq_GFbar}
\end{eqnarray}
with $\beta=1/T$ being the inverse temperature and $V$ the volume of each configuration.

\begin{figure}[t]
\centerline{\resizebox{.9\columnwidth}{!}{\includegraphics*{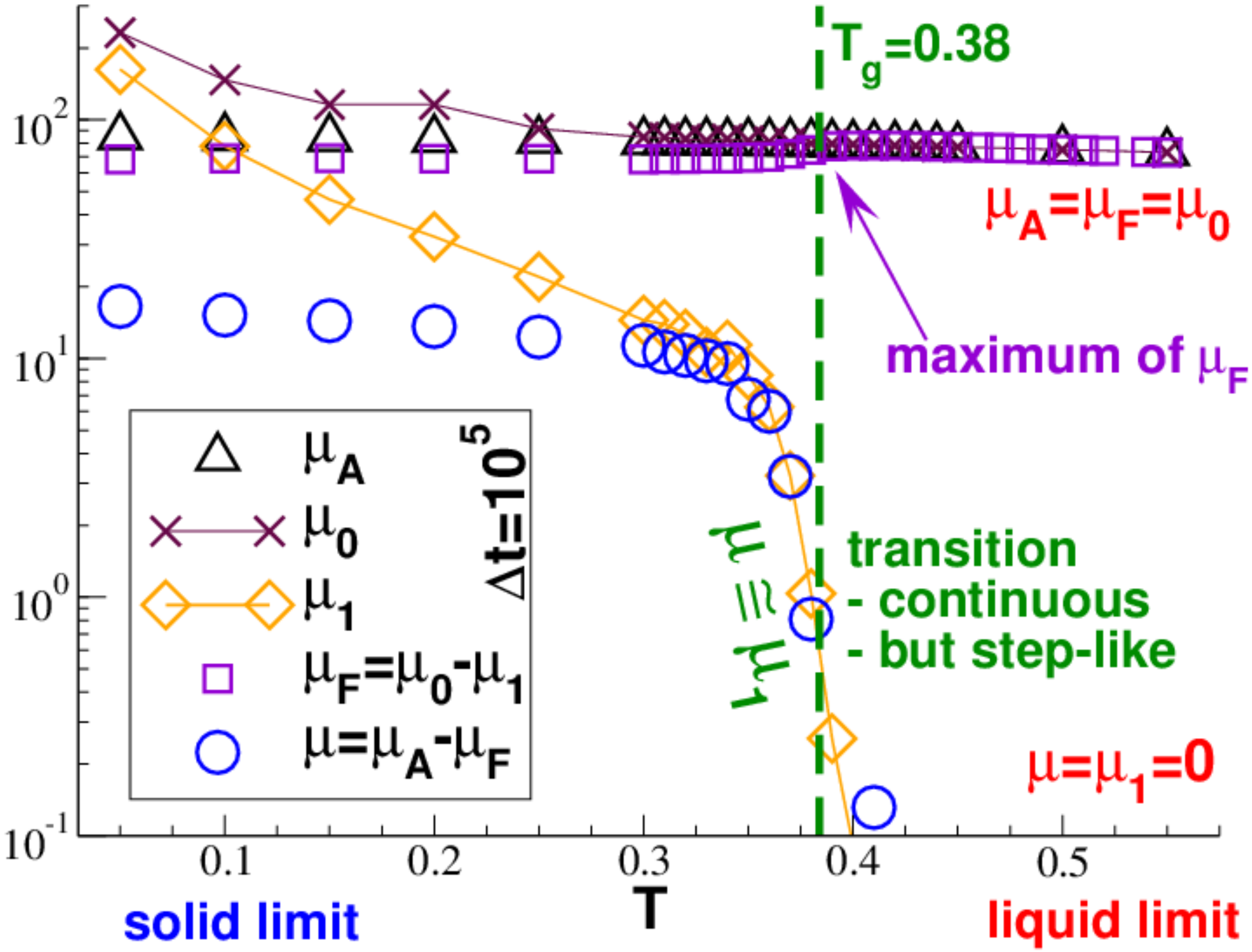}}}
\caption{First moments $\muA$, $\muFtwo$, $\muFone$, $\muF$ and $\GF$ {\em vs.} temperature $T$
using a half-logarithmic representation.
The data are obtained for $\tsamp=\tsampmax=10^5$.
With decreasing temperature $\GF \approx \muFone$ increases rapidly at $\Tglass$, but remains continuous.
Interestingly, $\muFtwo$ deviates from $\muA$ and $\muFone$ from $\GF$ below $\Tglass$.
\label{fig_mean}
}
\end{figure}

\paragraph*{Expectation values.}
The corresponding ensemble averages
$\muA \equiv \langle \muAbar \rangle$,
$\muFtwo \equiv \langle \muFtwobar \rangle$,
$\muFone \equiv \langle \muFonebar \rangle$,
$\muF \equiv \langle \muFbar \rangle$ and
$\GF \equiv \langle \GFbar \rangle$ are then obtained by averaging over the $m$
configurations and the three shear planes \cite{foot_ensaver}.
We have already presented the modulus $\GF(T)$ in the main panel of Fig.~\ref{fig_key} using a 
linear representation. Figure~\ref{fig_mean} presents $\GF(T)$ and its various contributions for 
$\tsamp=\tsampmax=10^5$ using half-logarithmic coordinates. 
As emphasized above, albeit $\GF=\muA-\muF$ increases rapidly below $\Tglass$, 
the data remain continuous in line with findings reported for colloidal glass-formers
\cite{Barrat88,WXP13,LXW16} using also the stress-fluctuation formula.
As one expects $\GF=\muFone=0$ in the liquid limit above $\Tglass$ and, hence, $\muF=\muFtwo=\muA$ 
\cite{WXP13,WXB15,ivan17a}. At variance to this, $\muF < \muA$ below $\Tglass$, i.e. the shear-stress 
fluctuations do not have sufficient time to fully explore the phase space. 
In agreement with Lutsko \cite{Lutsko88} and more recent studies 
\cite{WTBL02,Barrat06,WXP13,LXW16},
$\muF$ does not vanish for $T \to 0$, i.e. $\muA$ is only an upper bound of $\GF=\muA-\muF$.
We emphasize that while $\muF=\muFtwo-\muFone$ is more or less constant below $\Tglass$, 
its contributions $\muFtwo$ and $\muFone$ increase rapidly with decreasing $T$. 

\begin{figure}[t]
\centerline{\resizebox{.9\columnwidth}{!}{\includegraphics*{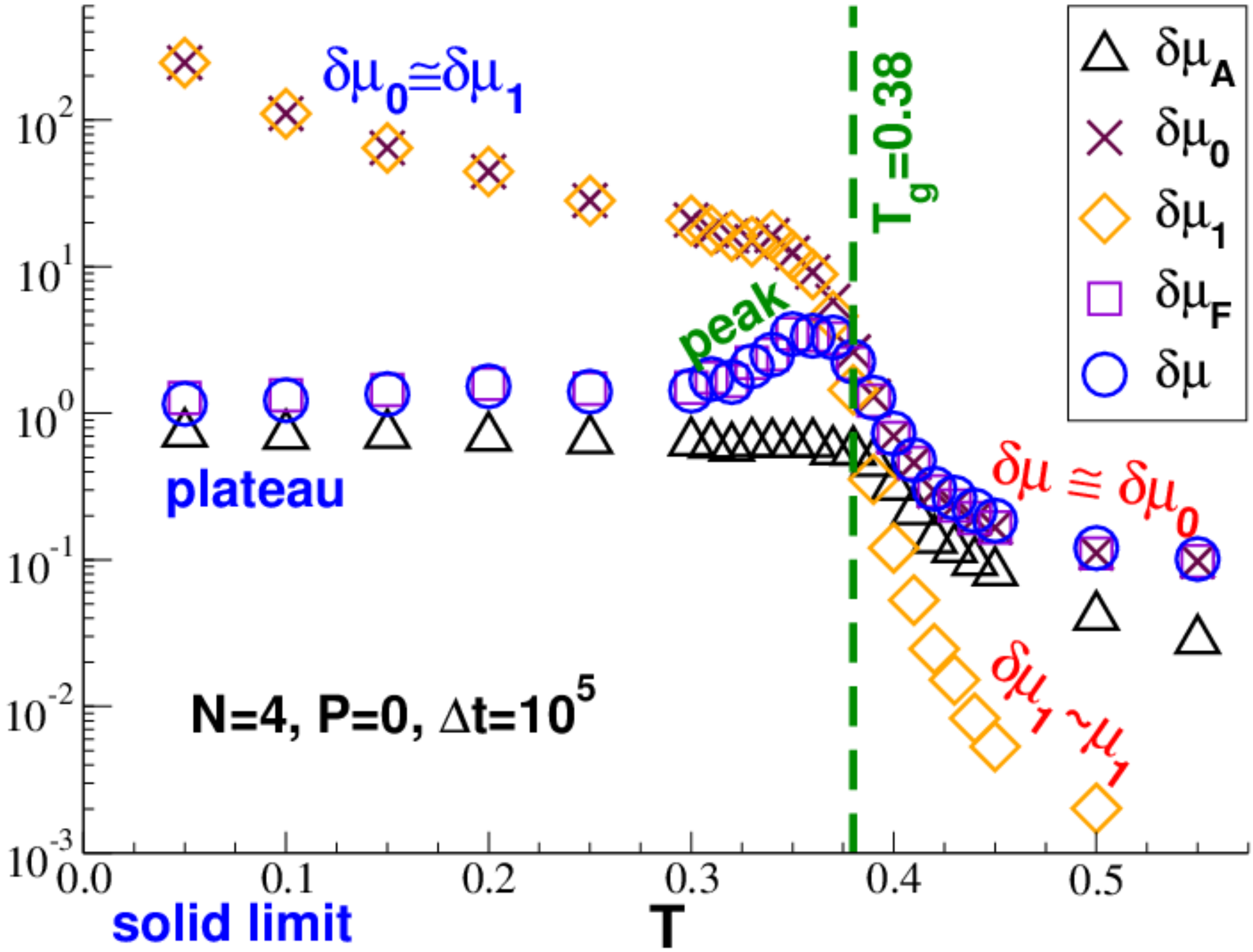}}}
\caption{Standard deviations $\dmuA$, $\dmuFtwo$, $\dmuFone$,
$\dmuF$ and $\dGF$ as a function of $T$.
$\dmuA$ is found to be small and $\dGF \approx \dmuF$ for all $T$.
Below $\Tglass$ the standard deviations $\dmuFone$ and $\dmuFtwo$ become rapidly similar and
orders of magnitude larger than $\dmuF$. This confirms the presence of strong frozen shear stresses.
\label{fig_fluctu}
}
\end{figure}

\paragraph*{Ensemble fluctuations.}
To characterize also the fluctuations between different configurations we take for 
various properties in addition the second moment over the ensemble.
As already seen in the inset of Fig.~\ref{fig_key}, we thus compute, e.g., the standard 
deviation of the shear modulus
$\dGF = \sqrt{\langle \GFbar^2 \rangle - \langle \GFbar \rangle^2}$ \cite{foot_ensaver}.
$\dGF$ is presented together with the corresponding standard deviations $\dmuA$, $\dmuFtwo$, 
$\dmuFone$ and $\dmuF$ in Fig.~\ref{fig_fluctu}.
As can be seen, $\dmuA$ is negligible and $\dGF \approx \dmuF$ for all $T$.
In the high-$T$ regime we find $\dGF \approx \dmuFtwo$ while $\dmuFone \approx \muFone$ 
vanishes much more rapidly.
In the opposite glass-limit $\dGF \approx \dmuF$ becomes orders of magnitude smaller than
$\dmuFtwo \approx \dmuFone$ \cite{ivan17a}.
The contributions $\muFtwobar$ and $\muFonebar$ of the difference $\muFbar=\muFtwobar-\muFonebar$ 
thus must be strongly correlated as one verifies using the corresponding correlation coefficient.
This is yet another manifestation of the strong frozen shear stresses which are
generated in each configuration while quenching the systems through the glass transition.

\begin{figure}[t]
\centerline{\resizebox{.9\columnwidth}{!}{\includegraphics*{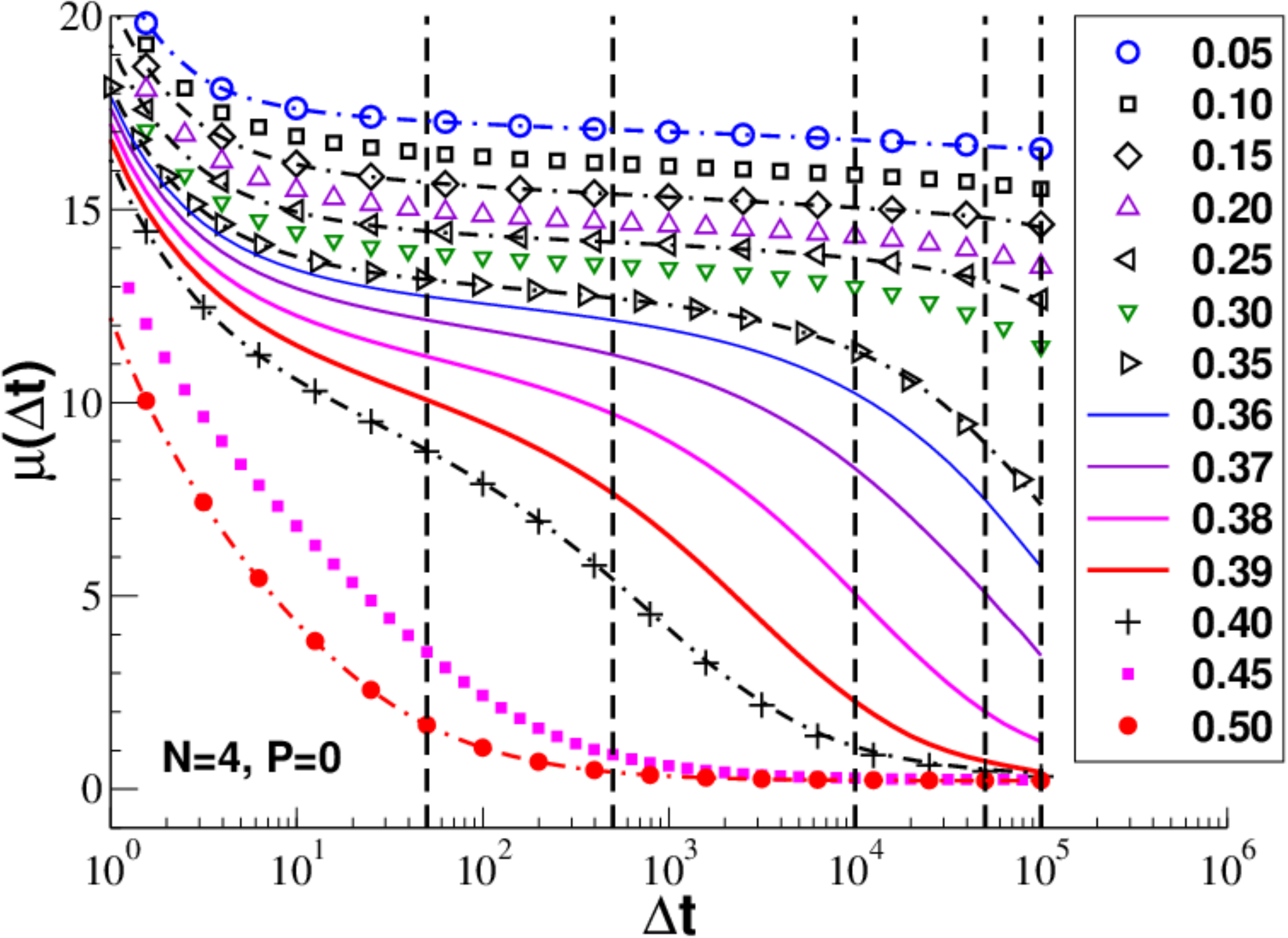}}}
\caption{Shear modulus $\GF$ as a function of sampling time $\tsamp$ for a broad range 
of $T$ as indicated in the figure. $\GF(\tsamp)$ decreases continuously with $\tsamp$.
Note that a smaller temperature increment $\Delta T =0.01$ is used around $\Tglass$ (solid lines) 
where $\GF(\tsamp;T)$ changes much more rapidly with $T$.
The dash-dotted lines are obtained using Eq.~(\ref{eq_GF_Gt}) by integrating
the directly measured shear-stress relaxation modulus $G(t)$.
The vertical lines mark the sampling times used in Fig.~\ref{fig_key}.
}
\label{fig_GF_dt}
\end{figure}

\paragraph*{$\tsamp$-dependence.}
%
We return now to the sampling time dependence shown in Fig.~\ref{fig_key}.
As expected from crystalline and amorphous solids \cite{LXW16,WXP13}
and permanent \cite{WXP13,WXB16} and transient \cite{WKC16} elastic networks,
the expectation values of the contributions $\muA$ and $\muFtwo$ to $\GF$ are strictly 
$\tsamp$-independent (not shown). 
This can be traced back to the fact that their time and ensemble averages commute \cite{WXB16,foot_muAmuFtwo}. 
This is strikingly different for $\muFone$, $\muF$ and $\GF$ for which this commutation relation does not hold.
%
%
As shown in Fig.~\ref{fig_GF_dt}, we focus here on the $\tsamp$-dependence of 
$\GF(\tsamp)=(\muA-\muFtwo)+\muFone(\tsamp)$.
Covering a broad range of temperatures we use subsets of length $\tsamp$ of the total
trajectories of length $\tsampmax$ stored.
It is seen that $\GF(\tsamp)$ decreases both monotonously and continuously with $\tsamp$.
The figure reveals that $\GF(\tsamp;T)$ decreases also monotonously and continuously with $T$.
A glance at Fig.~\ref{fig_GF_dt} shows that one expects the transition of $\GF(T)$ to get shifted
to lower $T$ and to become more step-like with increasing $\tsamp$ in agreement with Fig.~\ref{fig_key}. 
(Note that $\GF(\tsamp)$ increases for $T \to 0$ while its decay slows down.)
It is, however, impossible to reconcile the data with a jump-singularity at a {\em finite} $\tsamp$ and $T$.
It is neither possible to achieve a reasonable data collapse by shifting the data.

\begin{figure}[t]
\centerline{\resizebox{.9\columnwidth}{!}{\includegraphics*{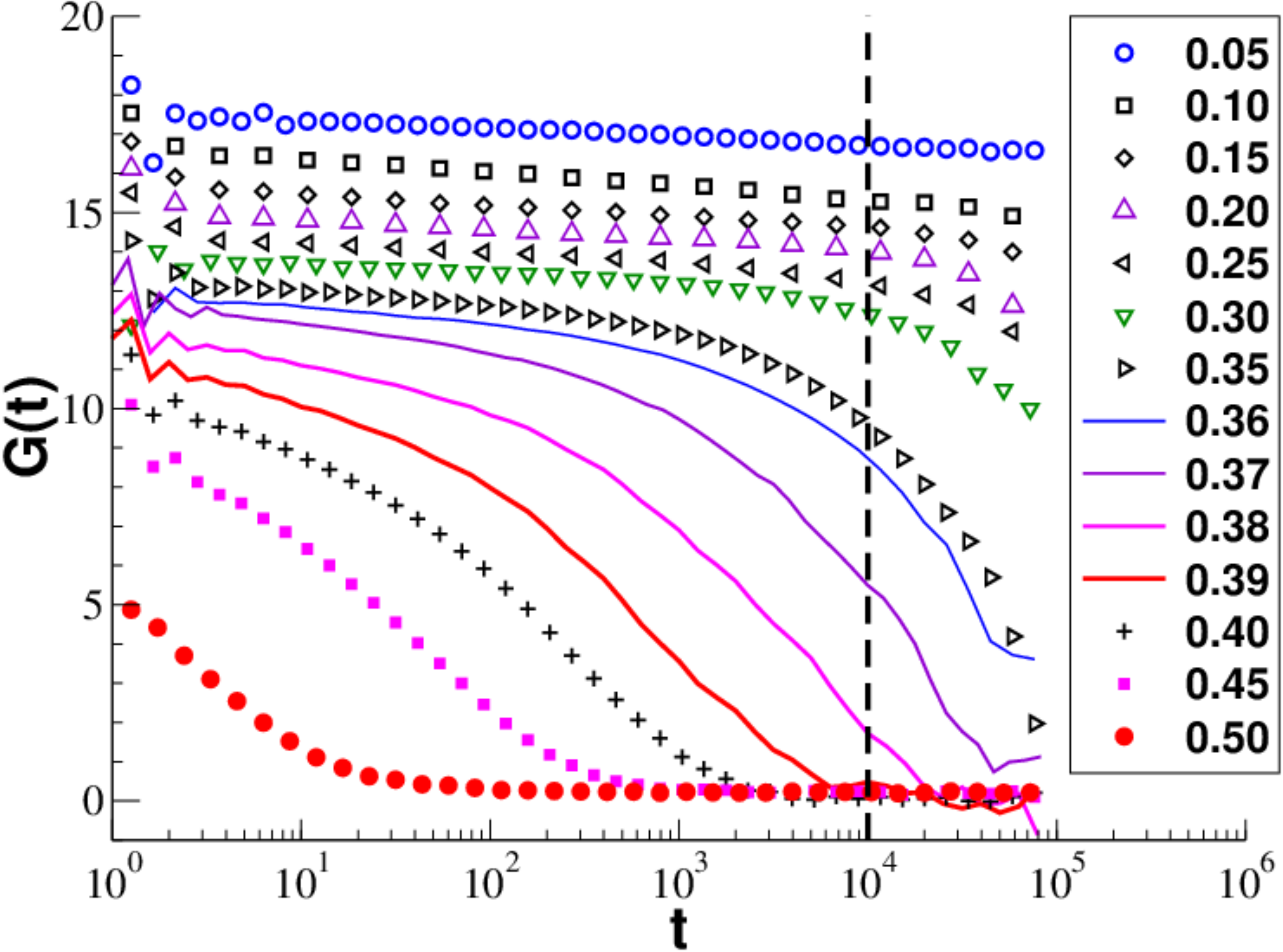}}}
\caption{Stress relaxation modulus $G(t)$ for a broad range of $T$.
The data are rather similar to the shear modulus $\GF(\tsamp)$ presented in Fig.~\ref{fig_GF_dt}.
The dashed vertical line marks the time $t=10^4$ used for $G(t)$ and $\delta G(t)$
in Fig.~\ref{fig_key}.
}
\label{fig_Gt}
\end{figure}

\paragraph*{Connection between $\GF(\tsamp)$ and $G(t)$.}
The systematic sampling time dependence of $\GF(\tsamp)$ shown in Fig.~\ref{fig_key} and Fig.~\ref{fig_GF_dt} 
can be understood from the generic sampling time dependence of time-averaged fluctuations 
\cite{LandauBinderBook}.
Assuming time-translational invariance $\GF(\tsamp)$ may be 
written as a weighted average \cite{WXB15,WXBB15,WKB15,WXB16,WKC16,ivan17a}
\begin{equation}
\GF(\tsamp) = \frac{2}{\tsamp^2} \int_0^{\tsamp} (\tsamp -t) \ G(t) \ \ddiff t,
\label{eq_GF_Gt}
\end{equation}
over the shear-relaxation modulus $G(t)$ \cite{foot_SM}. 
As shown in Fig.~\ref{fig_Gt}, we have computed $G(t)$ directly by means of the  
fluctuation-dissipation relation appropriate for canonical ensembles with quenched or sluggish 
shear stresses \cite{WXB15,WXB16,ivan17a,foot_SM}.
Having thus characterized the relaxation modulus $G(t)$, the numerical sum corresponding to Eq.~(\ref{eq_GF_Gt})
yields the thin dash-dotted lines indicated in Fig.~\ref{fig_GF_dt}. Being identical with 
the stress-fluctuation formula $\GF = \muA-\muF$ for {\em all} $T$, this confirms the assumed 
time-translational invariance. The $\tsamp$-dependence of $\GF$, $\muFone$ and $\muF$ is thus simply 
due to the upper boundary $\tsamp$ used to average $G(t)$. 
As one expects from Eq.~(\ref{eq_GF_Gt}) \cite{foot_SM}, the functional forms of $\GF(\tsamp)$ and $G(t)$ 
are rather similar, especially at low $T$.
Fixing a time $t$, say $t=10^4$ as indicated by the vertical dashed line in Fig.~\ref{fig_Gt}, 
allows to characterize the temperature dependence of the relaxation modulus $G(t)$ and its 
standard deviations $\delta G(t)$ (bold solid lines in Fig.~\ref{fig_key}).
Consistently with Eq.~(\ref{eq_GF_Gt}), a similar behavior is found as for $\GF(T)$ and $\dGF(T)$. 

\begin{figure}[t]
\centerline{\resizebox{.9\columnwidth}{!}{\includegraphics*{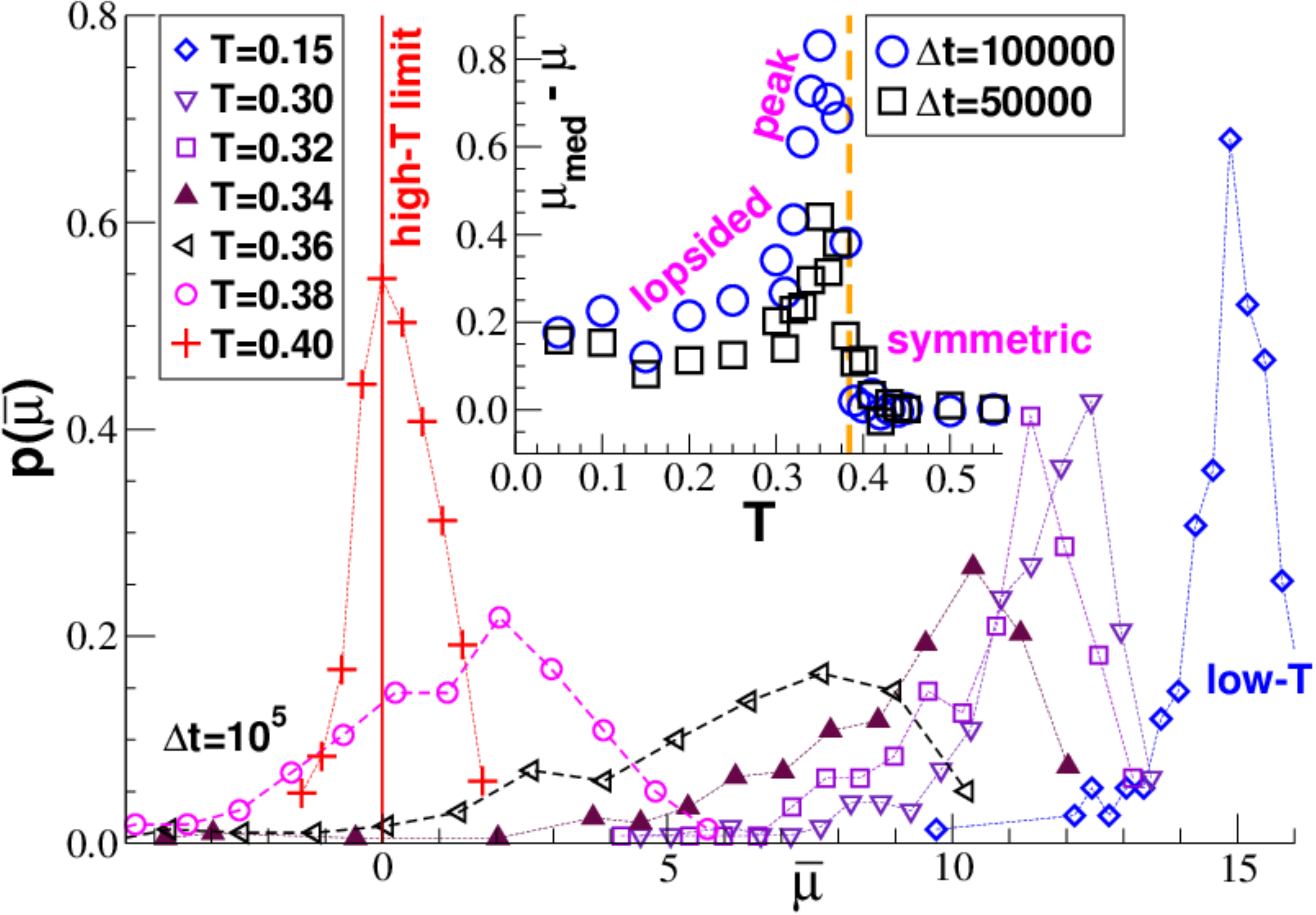}}}
\caption{Main panel:
Distribution $p(\GFbar)$ for $\tsamp=10^5$ for different $T$.
Inset: Difference $\GFmed-\GF$ of the median $\GFmed$ and the ensemble average $\GF$
{\em vs.} $T$ for two sampling times. The difference has a peak slightly below $\Tglass$ 
corresponding to very lopsided distributions.
}
\label{fig_histo}
\end{figure}

\paragraph*{Distribution of $\GFbar$.}
The striking peak of $\dGF$ below $\Tglass$ seen in Fig.~\ref{fig_key} begs for a more detailed
characterization of the distribution $p(\GFbar;T,\tsamp)$ of the time-averaged shear modulus $\GFbar$.
Focusing on our largest sampling time $\tsampmax$, the main panel of Fig.~\ref{fig_histo} presents
normalized histograms obtained using $3 \times m = 300$ measurements.
We emphasize that the histograms are unimodal for all $T$ and $\tsamp$ \cite{foot_histo_above}. 
The $T$-dependence of $\GF$ and $\dGF$ below $\Tglass$ seen in Fig.~\ref{fig_key}
is thus not due to, e.g., the superposition of two configuration populations representing
either solid states with finite $\GFbar$ and liquid states with $\GFbar \approx 0$.
The maximum $\GFmax$ of the (unimodal) distribution systematically shifts to 
higher values below $\Tglass$, in agreement with its first moment $\GF$ (Fig.~\ref{fig_key}),
while the distributions become systematically broader and more lopsided, 
i.e. liquid-like configurations with small $\GFbar$ remain relevant. 
The increase of $\dGF$ with sampling time $\tsamp$ seen in the inset of Fig.~\ref{fig_key} 
is due to the broadening of $p(\GFbar)$ caused by the growing weight of small-$\GFbar$ configurations (not shown).
For even smaller temperatures $T \ll \Tglass$, the distributions get again more focused around their maxima $\GFmax$ 
(as expected from Fig.~\ref{fig_key}) and less lopsided. That the large standard deviations 
and the asymmetry of the distributions are related is demonstrated
by comparing the first moment $\GF$ of the distribution, its median $\GFmed$ 
and its maximum $\GFmax$. One confirms that $0 < \GFmed - \GF < \GFmax -\GF$ below $\Tglass$ for all $\tsamp$.
As seen from the inset of Fig.~\ref{fig_histo}, $\GFmed-\GF$ has a peak similar to $\dGF$ 
becoming sharper with increasing $\tsamp$.

\paragraph*{Summary.}
We investigated by means of MD simulations a coarse-grained model for polymer glasses
characterizing its shear modulus $\GF$ using the stress-fluctuation formalism.
The observed $\tsamp$-dependence of $\GF$ (Fig.~\ref{fig_key}) 
and its contributions $\muFone$ and $\muF$ can be traced back to the finite time (time-averaged)
stress fluctuations need to explore the phase space which is perfectly described 
(Fig.~\ref{fig_GF_dt}) by the weighted integral over the shear-stress relaxation modulus $G(t)$ 
[Eq.~(\ref{eq_GF_Gt})].
The liquid-solid transition characterized by the ensemble-averaged $\GF(T)$ is continuous
for all sampling times $\tsamp$, but becomes sharper and thus better defined with increasing $\tsamp$ 
(Fig.~\ref{fig_key}). 
However, while the transition gets more step-like {\em on average}, increasingly strong fluctuations 
between different configurations underly the transition. The broad and lopsided distribution $p(\GFbar)$ 
below $\Tglass$ makes the prediction of the modulus $\GFbar$ of a single configuration elusive 
(Fig.~\ref{fig_histo}).

\paragraph*{Beyond the current study.}
While $\GF$ and its contributions $\muA$, $\muF$, $\muFtwo$ and $\muFone$ do not depend on the 
system size \cite{foot_SM},
this is more intricate for the corresponding standard deviations and must be addressed 
in the future following Ref.~\cite{Procaccia16}.  
Recent work on self-assembled networks \cite{WKC16} suggests
that $\dGF \approx \dmuF \sim 1/\sqrt{V}$ for $T \ll \Tglass$ (self-averaging),
while $\dGF \approx \dmuF \sim V^0$ around $\Tglass$ (lack of self-averaging).
In the latter limit long-ranged elastically interacting activated events 
are expected to dominate the plastic reorganizations of the particle contacts \cite{Barrat14b}.
From a broader vantage point it is no surprise that the lifting of the permutation
invariance of the liquid state below $\Tglass$ \cite{Alexander98} 
should lead to strong fluctuations between different configurations. 
The observation of strong frozen-in shear stresses 
(Figs.~\ref{fig_mean} and \ref{fig_fluctu}) is 
thus merely a demonstration of the broken symmetry. 
Analysis tools need to account for these frozen zero-wavevector stresses 
and theoretical approaches neglecting them are bound to miss the heart of the problem.
%
%

\vspace*{0.2cm} 
\begin{acknowledgments}
I.K. thanks the IRTG Soft Matter for financial support.
We are indebted to O. Benzerara (ICS, Strasbourg) and H.~Xu (Metz) for helpful discussions.
We thank the University of Strasbourg for a generous grant of cpu time through GENCI/EQUIP${@}$MESO.
\end{acknowledgments}


\begin{thebibliography}{43}
\bibitem[{foo({\natexlab{c}})}]{foot_muAmuFtwo}
\bibinfo{note}{$\muA$ and $\muFtwo$ are in this sense static observables. The
  observation that $\muA$ and $\muFtwo$ systematically deviate below $\Tglass$
  (Fig.~\ref{fig_mean}), thus implies that the glass transition of our polymer
  melts cannot be interpreted in terms of a purely dynamical theoretical
  framework.}
\bibitem[{foo({\natexlab{d}})}]{foot_histo_above}
\bibinfo{note}{For $T > \Tglass$ the distribution is Gaussian, centered at a
  maximum $\GFmax = 0$ and getting sharper with increasing sampling time
  $\tsamp$ (not shown).}
\end{thebibliography}

\end{document}